# Dark Energy Spectroscopic Instrument (DESI) Fiber Positioner Production

Daniela Leitner[*a], Jessica Aguilar[a], Jon Ameel[b], Robert Besuner[c], Todd Claybaugh[a], Henry Heetderks[c], Michael Schubnell[b], Jean-Paul Kneib[d], Joseph Silber[a], Gregory Tarlé[b], Curtis Weaverdyck[b], Kai Zhang[a]

[a]Lawrence Berkeley National Laboratory, 1 Cyclotron Rd, Berkeley, CA, USA 94720;
[b]University of Michigan, Randal Laboratory, 450 Church Street, Ann Arbor, MI, USA 48109;
[c] Space Sciences Laboratory (SSL), 7 Gauss Way, Berkeley, CA 94720
[d]École Polytechnique Fédérale de Lausanne, LSRO, Station 9, ME B3 465, CH-1015, Lausanne, CH, Switzerland

## ABSTRACT

The Dark Energy Spectroscopic Instrument (DESI) is under construction to measure the expansion history of the Universe using the Baryon Acoustic Oscillation technique.  The spectra of 35 million galaxies and quasars over 14000 sq deg will be measured during the life of the experiment.  A new prime focus corrector for the KPNO Mayall telescope will deliver light to 5000 fiber optic positioners. The fibers in turn feed ten broad-band spectrographs. We will describe the production and manufacturing processes developed for the 5000 fiber positioner robots mounted on the focal plane of the Mayall telescope.

## 1.  INTRODUCTION

The Dark Energy Spectroscopic Instrument (DESI) will perform an optical/near-infrared survey to measure the effect of dark energy on the expansion of the universe with a planned start of the observations in October of 2019. Over a 5 year time period DESI will obtain optical spectra for tens of millions of galaxies and quasars, constructing a 3-dimensional map spanning the nearby universe to 10 billion light years. DESI will be installed at the Kitt Peak National Observatory on the 4-meter Mayall telescope. The optical design of the Mayall is based on a prime-focus corrector. Light enters the telescope from an area on the sky and is reflected from the 4 meter mirror into a focal plane. As part of the telescope refurbishment, the existing corrector will be replaced with a new DESI corrector and a dynamic Focal Plane Assembly (FPA) which will provide fast repositioning of the 5000 science fibers. The in-situ repositioning capability enables the planned magnitude of the survey over the five-year observation period. This capability to rapidly adjust the 5000 fibers of the focal plane between each observation (about every 20 minutes) is a major advance in respect to previous instruments and is key to the number of targets that can be reached during the DESI survey. This requirement however places stringent specification on the positioners [1]. Each of the 5000 Fiber Positioners needs to be individually controlled by its own control board. Each fiber positioner has to consistently locate its fiber on target with an accuracy of better than 5 micrometer RMS. All 5000 fibers need to be in position in less than 120 seconds between exposures. In case this activity cannot be performed in parallel with other between-exposure activities, the positioner reconfiguration needs to be complete in 45 seconds. To accommodate the survey under all operating conditions of the Mayall Telescope, the fiber positioners have to be able to consistently perform at the specified positioner accuracy for more than 100,000 moves and a temperature range of -10 to +30 degrees Celsius. In order to maximize the field of view the positioners are densely packed with an average 10.4 mm center-to-center pitch which is less than their patrol disk of 12 mm. Consequently, during repositioning collisions with their close neighbors need to be avoided by the control algorithm. The push to minimize the space occupied by the fiber positioners and maximize the observation coverage is a design driver. Minimizing the positioner to a 10mm envelope has important consequences in terms of the required mechanical tolerances of the positioner parts and the assembly requirements (see section 3).

*DLeitner@lbl.gov; phone: 1-510-486-4503; Lawrence Berkeley National Laboratory, 1 Cyclotron Rd, Berkeley, CA, USA 94720

The original positioner design and assembly sequences were developed at the Lawrence Berkeley National Laboratory. The actual mass-production of the full robotic assemblies is carried out at the University of Michigan. The 107 micrometer diameter fiber assemblies are manufactured at the Lawrence Berkeley National Laboratory (LBNL). In addition, final integration of the positioner and the fibers is performed at LBNL. In this step the fibers are installed and precision aligned in the positioners at LBNL to ±15 micrometers. After this step, the integrated positioners are installed into the Focal Plane [1, 2, 3].

## 2. MECHANICAL DESIGN OF THE POSITIONER

The design is described in detail in previous proceedings [4]. Here we briefly summarize some key features that are important for the manufacturing process. A mechanical model and a picture of the completed fiber positioner are shown in Figure 1 together with the individual components.

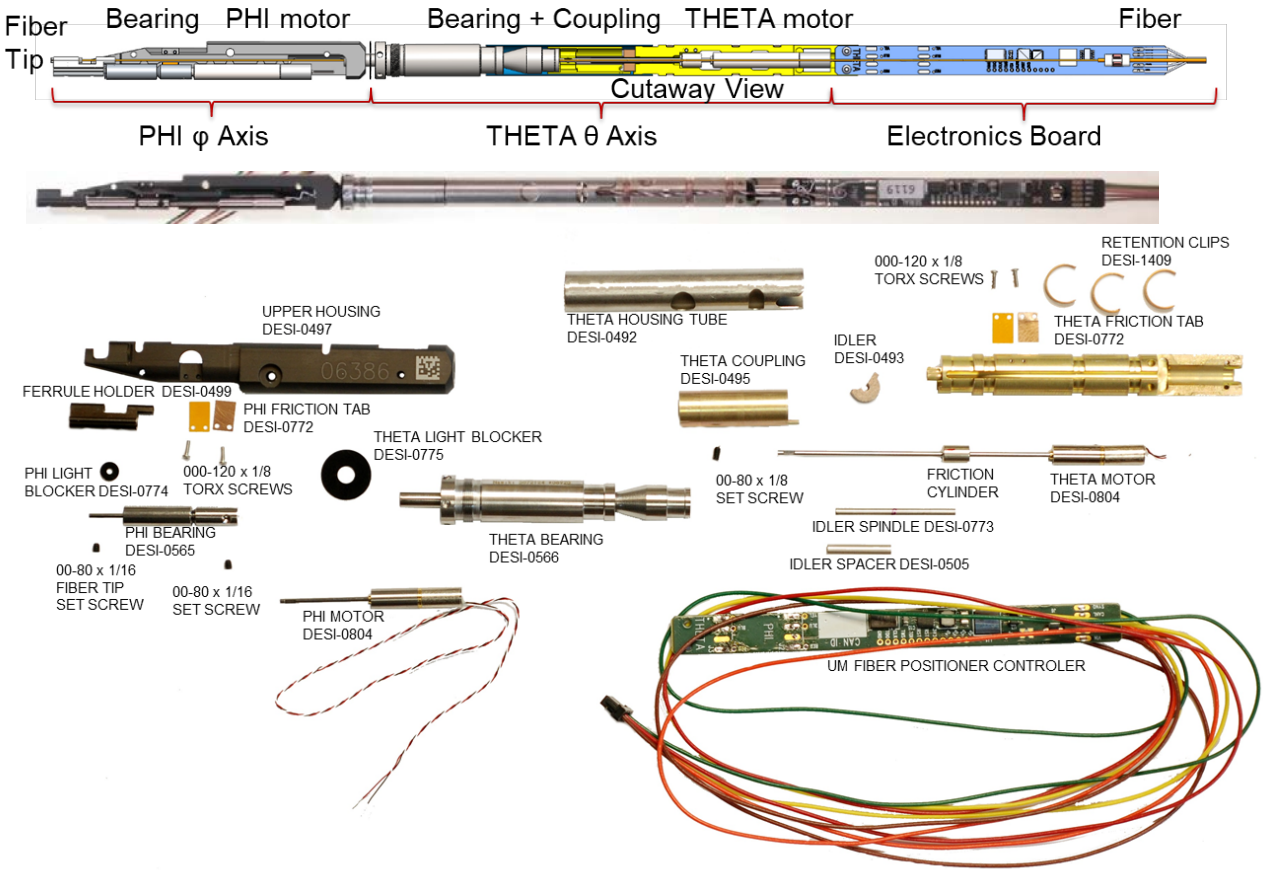

Figure 1. Fiber positioner CAD model, an actual positioner, and the individual components.

The fiber is positioned utilizing two 4 mm diameter DC brushless gear motors (manufactured by NAMIKI Precision Jewel Co., Japan). It is routed through the phi assembly, passes through the theta bearing and is then guided smoothly through the interior of the remaining theta assembly. The fiber routing is designed to stay outside of a minimum bend radius and such that the positioner can accommodate the full 360 degrees of rotation without damaging the fiber. Theta and phi axis hardstops prevent the positioner from rotating beyond this limit in case of a control software glitch. Finally, the positioner is designed to absorb stray light and reduce reflections - the upper housing that contains the phi motor

*DLeitner@lbl.gov; phone: 1-510-486-4503; Lawrence Berkeley National Laboratory, 1 Cyclotron Rd, Berkeley, CA, USA 94720

assembly is black anodized and any metal surface that is exposed to the focal plane is shielded by black Delrin light blockers.

The positioners have a eccentric axis kinematics (one on-axis theta motor, and one off-axis (eccentric) phi motor) which enables positioning of the fiber at any point within the 12 mm patrol disk [4]. The positioners do not utilize encoders and are controlled in open loop. The positioners use optical feedback from the Fiber View Camera located in the middle of the telescope mirror that images the back-illuminated fibers and reaches their target in iterative move sequences [4]. For each positioner calibration data such as the precise move radii for the theta and phi arms as well as the operational current (power) required to achieve the specified positioning accuracy are generated. The positioner location and address are mapped in the focal plane and stored in a database. Consequently, each positioner is controlled based on its unique calibration file to maximize light throughput and positioning accuracy. It also optimizes the required power to operate the fiber positioner array in the focal plane. Every positioner is extensively tested in a standard testing sequence that includes a 'burn-in phase' of 5000+ moves to eliminate infant mortality of the assembly [5,6]. In addition, extended lifetime tests (see section 4) were performed on subsets of the production positioners, and whenever design or production changes were implemented, to ensure that the positioners can meet the specified lifetime requirement of 100k+ moves in the instrument. As of April 2018, we have fabricated more than 4500 positioners with an overall 80% mechanical assembly yield and 95% passing the final qualifying positioning accuracy test [5].

## 3. MANUFACTURING OF FIBER POSITIONER

During an extensive pre-production phase and the early production phase, the assembly processes and workflows were optimized for the required peak production rate of 50 assemblies per day (the average production rate is 180 robots per week). Stringent quality assurance steps were developed for each step of the production line to ensure consistent quality of the robots during the 12 months of production. A custom database with an iOS touch screen data entry interface was developed at the University of Michigan to manage the inventory and location of more than 200,000 parts. The database is key to manage the parts logistics from the international suppliers to the manufacturing location at the University of Michigan.

Including spares production, a total of 6000 functional positioners are required to complete the DESI project. With the current production rate of more than 640 positioners per month we project that the production will be completed in the Summer of 2018. Getting to this rate took significant effort. Elements that support the actual assembly line are

- A flexible integrated database to handle data throughout the process (part receipt -> shipping), providing data summaries to track quality and progress
- Active vendor management to ensure parts production can support the production workflow
- Stringent control of parts storage, QA, inspection tooling, and cleaning
- Stringent quality control of incoming parts and tooling
- Assembly instructions and assembly quality control procedures
- Development of a final assembly test workflow and procedures including a consistent grading system
- Utilizing an issue list that tracks manufacturing issues and failures
- Development of assembly handling, storage, and shipping procedures
- Development of acceptance procedures at LBNL to qualify a positioner for focal plane integration

The following sections summarize these developments in more detail. Key findings and challenges encountered are discussed.

### 3.1 PREPRODUCTION

After an extended R&D phase during which several design alternatives were explored, a small number of positioners (30+) were fabricated and a subset was successfully tested on-sky as part of a small-scale prototype instrument (ProtoDESI, [7]). Following the successful R&D phase, parts for 1000 positioners were procured for an extensive pre-production phase of about 10 months. During this phase the initial set of production tooling and an assembly sequence were developed and consequently optimized for the parallel production of 50 positioners per day.

*DLeitner@lbl.gov; phone: 1-510-486-4503; Lawrence Berkeley National Laboratory, 1 Cyclotron Rd, Berkeley, CA, USA 94720

Key lessons learned from the preproduction phase were

- Due to the stringent alignment requirements of the positioner assembly, tooling accuracy needs to be carefully assessed and optimized and regularly inspected for wear and glue residue
- The stringent tolerance requirements for the DESI positioner parts require QA screening of each part throughout the production line (even if the same supplier had already delivered thousands of parts successfully)
- In addition to the design effort, design changes require an extensive testing period at sufficient statistics (lifetime tests (>3 weeks), thermal tests (>2 weeks), procedure and manufacturing development (>2 weeks), time to fabricate and procure new tools and parts (2-3 months), training of technicians (1-2 weeks).
- Depending on the complexity of the design change, an evaluation will take 3-6 months before the changes are fully implemented in the production flow. Therefore, production line changes must be developed in parallel to the ongoing production and carefully evaluated for overall cost and performance benefit.
- Interrupting a parts supplier production line can lead to long delays in parts deliveries and quality issues on high precision parts (DESI parts have tolerances in the order of a few microns).
- Separation of the assembly line technicians and quality control is crucial to allow for fast feedback and improvements of the assembly lines.
- Preproduction is necessary to establish reliable production yields to manage parts procurements

The DESI positioner design is aggressively optimized to minimize the final envelope of the assembly. Therefore, it was critical to develop quality control check points for each assembly step to optimize the tooling for alignment. However, for several assemblies it is not possible to directly check alignment of the parts to each other after an assembly step. In addition, all assemblies rely on glue joints and cannot be disassembled once completed. In these cases, we are relying on the tooling accuracy and assembly procedure to ensure that parts and subassemblies are adequately aligned. We found that high resolution CT (Computerized Tomography) x-ray scans are an invaluable tool to analyze issues encountered in these cases. As an example, Figure 2 shows such a high-resolution 3D scan performed on one of our completed positioner assemblies. This analysis was done in collaboration with the Los Alamos National Laboratory.

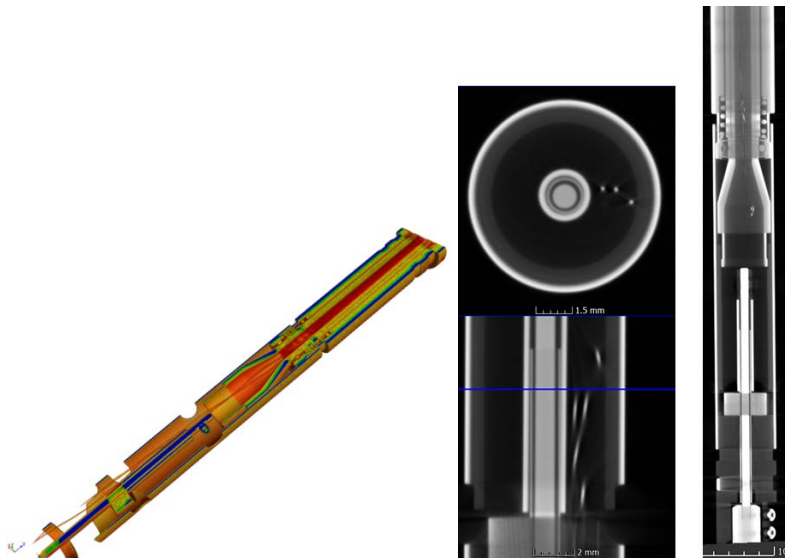

Figure 2. Example of a 3D CT scan of a completed (well aligned) positioner assembly performed at the Los Alamos National Laboratory. Shown is a cross section of the Theta coupling and Theta Bearing. The cable routing of the PHI motor cables can be seen in the theta coupling on the right – following the image our assembly procedure was modified and required that all motor cables are breaded before they are installed into the positioner to better control the cable and fiber routing.

*DLeitner@lbl.gov; phone: 1-510-486-4503; Lawrence Berkeley National Laboratory, 1 Cyclotron Rd, Berkeley, CA, USA 94720

This technique was also utilized to diagnose an intermittent issue we encountered with the 4 mm DC motors. The cross-sectional scans of the motor gearbox revealed that our assembly procedures risked compression of the planar gearbox [6]. We were able to eliminate this risk by implementing a minor change to our assembly procedure and tooling. In addition, we implemented additional quality control steps and are testing the motors upon receiving, after they are assembled into the subassemblies, and before the assembly is handed off to integration with the electronics board.

An important part of the pre-production phase was establishing touch times (the time an assembly is actually being worked on) to understand the workforce required to finish production on schedule. In addition, in order to manage the production flow and parts supply chain, firm production yields must be established. Table 1 summarizes the achieved production yields, and assembly times. A picture of each assembly and the assembly flow is shown in Figure 4.

|  | Yields | Time/assembly[min]* |
|---|---|---|
| ASSY 1 | 96% | 5 |
| ASSY 2 | 98% | 3 |
| ASSY 2A | 99% | 3 |
| ASSY 3 | 95% | 15 |
| ASSY 4A | 98% | 10 |
| ASSY5A | 95% | 5 |
| ASSY6 | 100% | 5 |
| ASSY7 | 99% | 10 |
| Mechanical Yield | 80% |  |
| XY | 97% | 24hrs** |
| Final Acceptance | 96% | 15 |

*assembly time only, does not include incoming parts QA or cleaning
**test stands can accommodate 39 positioners simultaneously, typically 25 positioners are tested in one set-up.

Table 1. Achieved average production yields and average fabrication times for the DESI positioner.

### 3.2 PRODUCTION DATABASE, MANUFACTURING CONTROLS, AND QUALITY CONTROL

Quality control of parts and sub-assemblies require standard metrology tools, microscopes for visual inspection, a custom motor driver, and a custom axial load test for the phi motor bearings. Dial indicator, granite surfaces and V-blocks are utilized to verify that the part runout is within specifications. Visual inspections are performed on each part by naked eye or utilizing a compact microscope to inspect for burrs and other manufacturing imperfection. If necessary, parts are reworked (polished, burrs removed, or parts re-tapped) to comply with our stringent inspection requirements. Visual inspection of assemblies, microscope inspections, and manual turn/twist of bearing post to feel for friction are performed for each part. Visual inspection is performed to verify epoxy amounts, thread locker amounts, and screw engagement. As mentioned above, each motor is tested several times during assembly to ensure that only performing subassemblies are moved forward to the next step. By sorting out failing motor assemblies early, only passing positioners are tested in the final 24-hour positioning test saving time and improving the overall throughput rate. Clear responsibilities have been established for several critical junction points in the production line including inventory management, workforce management, shipping, and quality control.

#### 3.2.1 PARTS MANAGEMENT

A total of 29 parts (see Figure 1) are required to assemble a positioner. Some fasteners and machined parts are used more than once in the assembly resulting in 22 unique parts. Considering the production yields, it was necessary to purchase parts for more than 7,000 positioners. Consequently, more than 200,000 individual parts had to be managed.

*DLeitner@lbl.gov; phone: 1-510-486-4503; Lawrence Berkeley National Laboratory, 1 Cyclotron Rd, Berkeley, CA, USA 94720

Tracking all parts is essential for manufacturing, project planning, and budgeting. Tracking a high volume of parts can be very time consuming if not managed from the beginning with discipline. Having a reliable part count confirms that enough parts are on hand to complete the required number of sub-assemblies/positioners and allows for work planning. A few straight forward work practices made this daunting task manageable.

Designated areas were created to handle incoming parts. The storage area needs to consider the actual workflow and the time it takes to move parts from receiving through incoming inspection to production. Simple actions such as keeping parts sealed and manually labeling the part bundles with critical information when they are received (such as part number, description, lot, date received, and quantity) reduce database entry errors and ease audits of the electronic database. Handling thousands of parts makes periodic internal part audits necessary to confirm the database balances match the actual balances. Reliable part counts are critical to keep the production line well stocked.

In case of DESI, all parts are managed through a SQL database that has an easy accessible web interface via Google sheets to readout quantities and assess part needs. The parts are organized by serial number or for non-serialized parts by lots. As parts are received they undergo the required QA checks and are then consequently moved from 'Received' to 'Available' or 'Rejected/Fail'. As assemblies are built electronic travelers are used which move parts in the database from 'Available' to 'Consumed'. At the same time the serial numbers or lot numbers for the assembly are recorded for each individual positioner. This allows tracking of QA issues back to a particular batch of parts and/or assembly steps.

### 3.2.2 MANUFACTURING CONTROL

Each subassembly process is recorded in an electronic traveler system using a custom iPad app that interfaces with the parts SQL database. The traveler has two purposes - it tracks that standard assembly procedures are followed (checklists) and records time, date, the technician performing the work, part serial numbers, and part lot numbers for quality control. Only if the final QA check is completed satisfactorily and signed off, the subassembly is moved to an 'Available' status for subsequent assemblies. This step is performed separately from the production lines. It ensures that failures are identified as early as possible in the process and removed from the assembly line if necessary. In the preproduction phase the travelers were initially developed using google sheets and paper travelers to provide specification for the data input, management and display to be implemented in the electronic database. In this way the programming time was reduced and the technicians performing the work could help optimizing the processing and were well accustomed to the traveler.

The production areas were functionally optimized to support the tasks required in the assembly line in the same way as the parts management areas. At an assembly flow of more than 200 subassemblies a day within a small cleanroom, it is important to have only the parts necessary for the next step in each area. The individual areas have designated spaces for 'Complete' sub-assemblies. These then move on to a final QA check and are next moved to areas labeled 'Available' if they pass QA inspection, 'Needs Review' if further investigation is needed, or 'Fail'. Our assembly procedures require that each part is plasma cleaned in the epoxy bond zones within one hour before assembly to improve the bonding strengths of the joint. All epoxy joints include witness samples for each epoxy batch mixed to ensure that the two-part epoxy was mixed correctly and that the glue joint was sufficiently cured before the next assembly step.
Critical screening steps that repeat some inspections already performed in the initial parts QA are part of the assembly checklist and have improved the production yields. 'Failing' parts or 'Needs Review' parts are set aside and removed from the assembly areas daily to reduce clutter and errors. After subassemblies pass QA (electronic sign-off step), they are set to 'Available' status and only then the electronic traveler allows these parts and subassemblies to be used in subsequent assemblies. This gating step was implemented during preproduction and has significantly improved our overall yields and assembly quality.
By optimizing workspaces, grouping tasks appropriately (reducing unnecessary movements) and applying stringent quality controls, we were able to decrease the mechanical assembly time for a positioner from more than two hours to about 45 minutes and improve the production yield from less than 50% to over 80%.

### 3.3 POSITIONER ASSEMBLY WORKFLOW

The assembly workflow is shown in Figure 4. A total of seven subassemblies are combined to fabricate one positioner. The most expensive parts of the assembly are the bearings, the motors, and their associated housing parts (Upper

*DLeitner@lbl.gov; phone: 1-510-486-4503; Lawrence Berkeley National Laboratory, 1 Cyclotron Rd, Berkeley, CA, USA 94720

Housing and Aft Cap). Therefore, the workflow was adjusted to individually test each axis assembly (theta and phi) separately before they are combined into the final positioner. This strategy minimizes the loss of expensive parts.

Our production flow allows for a peak production of 250 assemblies per day. However, if the part supply chain is not steady due to delays by vendors as this has been the case for our project, we have sufficient tooling in hand to shift production to fabricate a particular subassembly in advance. This strategy keeps the production flow steady and improves the workforce management.

A particular challenge for the DESI workflow management is the large utilization of undergraduate students for the production line. Managing changing semester schedules, semester breaks, and testing periods can be challenging, and it is therefore critical to balance student labor with full time technician and engineering support.

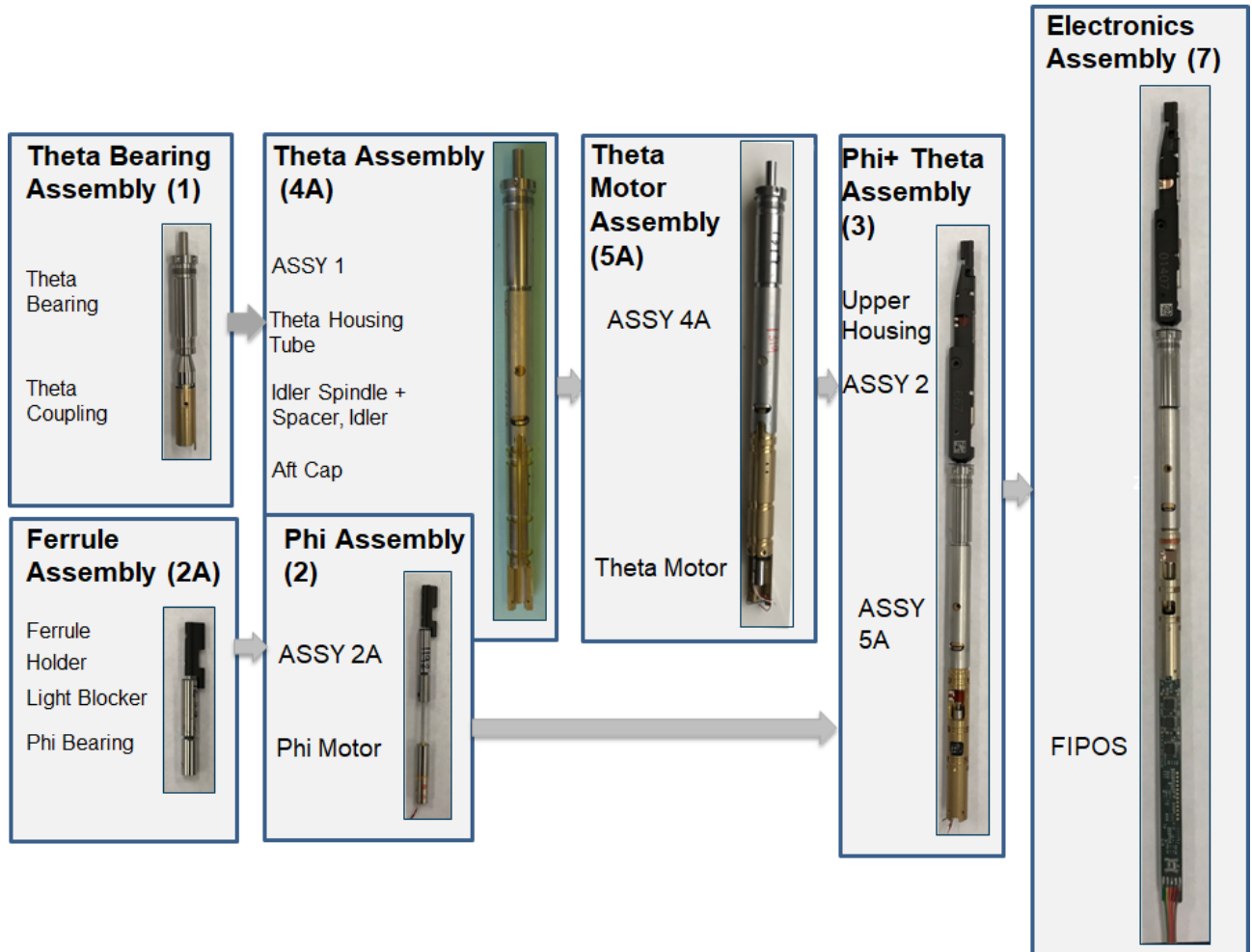

Figure 3 The assembly workflow as used in the production line (sub-assembly breakdown)

## 4. EXTENDED LIFETIME ANALYSES

In addition to the standard 24-hour burn-in test that includes the final grading of the positioner [5], a number of additional performance and engineering tests were performed during preproduction and production. This included extensive testing under various operating conditions such as the temperature range, operating angle, cooling fan positions, and air speeds. These tests are described in more detail in [5, 6].

*DLeitner@lbl.gov; phone: 1-510-486-4503; Lawrence Berkeley National Laboratory, 1 Cyclotron Rd, Berkeley, CA, USA 94720

In addition, extended lifetime tests were performed on a subset of positioners to ensure that the positioners can meet the requirement to complete 100,000+ moves during the five-year survey. These tests are time consuming, but invaluable to verify production quality. For example, during preproduction the extended lifetime helped identify and eliminate a problematic design requirement for our motors. To meet the overly stringent runout tolerance specified for the DC motors, the vendor added an additional bearing assembly to the motor which over constrained the drive shaft. While these positioners had initially excellent performances, extended lifetime tests showed that after 50,000 moves over 80% of the positioners failed the positioning accuracy test. We believe that these failures were due to wear out of the bearing assemblies due to the loading caused by misalignments in our assemblies. Once the additional bearing assemblies were removed the more compliant motor shaft could better compensate minor alignment errors of the mechanical assembly. Consequently, all the production positioners are built with the more compliant motor assemblies.

A statistical analysis was performed on 78 production positioners that were tested between 200 thousand moves (twice the lifetime move requirement) and 1.6 million moves (more than 16 time the lifetime move requirement). A large portion of positioners (91%) never failed during the tests and were only removed from the test set-up due to time constraints (resulting in censored test data). Two set of 12 positioners were run over a several months period to achieve more than a million moves after which the tests were terminated. To asses and project the lifetime expectancy of positioners in the petal a Weibull lifetime analysis was performed [8, 9]. The results are shown in Figure 4 and numerically summarized in Table 2. The results indicate that 98% of positioners will continue to perform after the DESI five-year survey is completed which easily exceeds the science requirement of 90% survival after 100,000 moves.

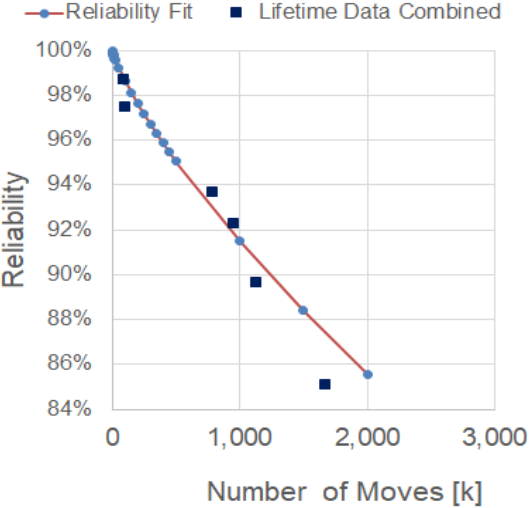

| Survival | kMoves |
|---|---|
| 99.5% | 29 |
| 99% | 68 |
| 98% | 161 |
| 95% | 508 |
| 90% | 1,234 |

Table 2. The numerical results of the best fit from Figure 4.

Figure 4. Summary of the lifetime analyses and best fit to the data.

## 5. SUMMARY

The DESI positioner production line development is mature, and the positioner assembly is well advanced. As of April 2018, we have completed more than 4500 production positioners and expect to finish production by mid-summer this year. The measured positioner performance is excellent; more than 95% of our positioner achieve or exceed the required positioning accuracy. Key lessons learned during preproduction and production are summarized in this paper to help plan future project of similar size and scale. The positioners are currently integrated into the DESI focal plane subassemblies for integrated testing and final assembly. Once completed the focal plane will then be shipped to the Mayall telescope for final integration and start of commission in 2019.

*DLeitner@lbl.gov; phone: 1-510-486-4503; Lawrence Berkeley National Laboratory, 1 Cyclotron Rd, Berkeley, CA, USA 94720

## 6. ACKNOWLEDGEMENTS

This research is supported by the Director, Office of Science, Office of High Energy Physics of the U.S. Department of Energy under Contract No. DE–AC02–05CH1123, and by the National Energy Research Scientific Computing Center, a DOE Office of Science User Facility under the same contract; additional support for DESI is provided by the U.S. National Science Foundation, Division of Astronomical Sciences under Contract No. AST-0950945 to the National Optical Astronomy Observatory; the Science and Technologies Facilities Council of the United Kingdom; the Gordon and Betty Moore Foundation; the Heising-Simons Foundation; the National Council of Science and Technology of Mexico, and by the DESI Member Institutions.  The authors are honored to be permitted to conduct astronomical research on Iolkam Du'ag (Kitt Peak), a mountain with particular significance to the Tohono O'odham Nation.

## 7. REFERENCES

[1]  DESI Collaboration, DESI final design report, Mar. 2016 http://desi.lbl.gov
[2]  Silber, J.H. et al., Design of the DESI Focal Plane System, Proc. SPIE, 10702-311 (2018).
[3]  Claybaugh, T.M. et al., Design of the DESI Focal Plane System, Proc. Proc. SPIE, 10706-217, (2018).
[4]  Schubnell, M. et al., Design of the DESI Focal Plane System, Proc. SPIE  9908,  doi: 10.1117/12.2233370 (2016).
[5]  Schubnell, M. et al., Design of the DESI Focal Plane System, Proc. SPIE, 10706-79 (2018).
[6]  K. Zhang et al., Design of the DESI Focal Plane System, Proc. SPIE 10706-161, (2018).
[7]  P. Fagrelius et al., ProtoDESI: Risk Reduction Experiment for the Dark Energy Spectroscopic Instrument, Proc. SPIE 9908-296, (2016).
[8]  Lawless, Jerald F., Statistical models and methods for lifetime data, Hoboken, N.J. : Wiley-Interscience, c2003.ISBN: 0471372153, (1944)
[9]  http://www.mathpages.com/home/kmath122/kmath122.html

*DLeitner@lbl.gov; phone: 1-510-486-4503; Lawrence Berkeley National Laboratory, 1 Cyclotron Rd, Berkeley, CA, USA 94720